\documentclass[journal]{IEEEtran}

\usepackage{xcolor}
\usepackage{amssymb}
\usepackage{amsmath}
\usepackage{algorithmic}
\usepackage{graphicx}

\begin{document}
\renewcommand{\thefootnote}{}
\title{Sequential HW-Aware Precoding: Over-the-air cancellation of HWI in Downlink Cell-Free Massive MIMO with Serial Fronthaul}
%
%
%

\author{Antoine~Durant,~\IEEEmembership{Member,~IEEE,}
        Asma~Mabrouk,~\IEEEmembership{Member,~IEEE,}
        and~Rafik~Zayani,~\IEEEmembership{Senior Member,~IEEE}}

\markboth{}%
{}

\maketitle

\begin{abstract}

This paper addresses the critical challenge of mitigating hardware impairments (HWIs) in downlink cell-free massive MIMO (CF-mMIMO) networks while ensuring computational scalability. We propose a novel sequential hardware-aware (HW-aware) precoding technique that leverages the serial fronthaul topology to perform over-the-air HWI cancellation. This approach involves sequentially exchanging approximated user-perceived distortion information among successive access points (APs) for over-the-air HWI mitigation. Each AP independently computes its spatial multiplexing weights and transmits signals that counteract the distortions introduced by the preceding AP. We develop a problem formulation and present a closed-form solution for this method. For performance evaluation, we study two reference methods taken either from centralized massive MIMO literature (Tone Reservation [TR]) or
tailored for CF-mMIMO networks (PAPR-aware precoding), both focusing on reducing the PAPR of OFDM signals in the downlink. Results indicate that the sequential HW-aware approach achieves a substantial increase in spectral efficiency (SE) in high-distortion scenarios, with an average SE increase factor of 1.8 under severe distortions. Additionally, the proposed method, which is executed locally, demonstrates better scalability, achieving a reduction of up to 40\% and 72\% in the total number of complex multiplications compared to the PAPR-aware and TR approaches, respectively. Finally, the sequential HW-aware precoder offers high performance even when applied to cost-effective APs with few antennas, presenting a promising and practical solution for HWI compensation in CF-mMIMO systems with serial fronthaul.

\end{abstract}


\IEEEpeerreviewmaketitle

\section{Introduction}
\footnotetext{Antoine Durant and Asma Mabrouk are with the CEA-Leti in Grenboble, France (emails:antoine.durant@cea.fr and asma.mabrouk@cea.fr). Rafik Zayani is with the IETR laboratory at the University of Rennes, France (email: rafik.zayani@univ-rennes.fr.}
The growing demand for wireless data has driven the evolution of communication systems towards higher capacity and energy efficiency. Conventional approaches such as cell-densification and co-located massive MIMO (multiple-input multiple output) are known to suffer from inherent limitations, including severe inter-cell interference and limited macro-diversity, which restrict their potential to scale efficiently~\cite{Lozano2013}. To address these challenges, a new paradigm termed Cell-Free massive MIMO (CF-mMIMO) has emerged, which has garnered significant attention in recent years~\cite{Chen2021}. Unlike traditional cellular networks, CF-mMIMO eliminates cell boundaries, with users being jointly served by a multitude of spatially distributed APs, providing enhanced spatial diversity and interference management. This distributed architecture offers numerous advantages, such as improved spectral efficiency and uniform service quality across the network, regardless of a user's location.

However, CF-mMIMO comes with its own set of challenges. The need to share large amounts of data, including user information, channel state information (CSI), and multi-user precoding vectors among the APs, introduces significant demands on the backhaul and fronthaul links. The sheer scale of data exchange can create scalability bottlenecks, as maintaining full cooperation among all APs is not feasible for large networks. Various strategies have been proposed to mitigate these scalability issues. For instance, local precoding schemes, where the multi-user precoders are computed locally at each AP without the need for extensive CSI exchange, have been shown to reduce backhaul traffic while maintaining a limited performance degradation compared to full cooperation scenarios~\cite{Interdonato2020, Bjoernson2020}. Additionally, innovative network topologies such as serial fronthaul architectures and radio stripes have been proposed to reduce the complexity of inter-AP communication. In such setups, advanced precoding strategies like compute-and-forward or Team-MMSE precoding, where cooperation is limited to adjacent APs, have demonstrated promising results in improving system performance with reduced signaling overhead~\cite{Shaik2020, Miretti2022}. However, these works consider perfect hardware in the APs and are based on a network-level power constraint rather than per-AP. Hence, their computational complexity is proportional to the number of AP in the network, which leads to scalability issues .

A further challenge in realizing large-scale CF-mMIMO systems lies in the hardware implementation. Given the extensive number of radio-frequency (RF) elements required, the use of low-cost and energy-efficient components is essential for economic viability. However, these cost-effective components typically introduce additional signal distortion, collectively referred to as hardware impairments (HWI). Such impairments have been extensively studied in the context of CF-mMIMO systems architectures~\cite{Zhang2018}. Among the key contributors to HWI are distortions caused by the power amplifiers (PAs). The non-linearities in low-cost PAs, including non-linear compression and clipping, introduce significant additive distortion noise, which can severely degrade the system performance~\cite{Mokhtari2021}.

To mitigate these hardware-induced distortions, two critical factors can be tailored specifically to minimize the distortions when designing CF-mMIMO systems based on low-cost PAs. First, the peak-to-average power ratio (PAPR) can be reduced to enhance energy efficiency, as high PAPR forces PAs to operate inefficiently with larger power back-offs. Second, it is essential to reduce distortion caused by non-linear effects in the PA, such as compression and clipping, which can otherwise limit system capacity and spectral efficiency. Advances in signal processing techniques offer potential solutions, enabling more effective utilization of the available hardware, thus enhancing spectral efficiency while minimizing the impact of HWIs. Techniques such as digital predistortion ~\cite{Kang1999} and PAPR reduction ~\cite{Wulich1999, Zayani2019} can help optimize the performance of hardware-constrained systems and are crucial to offer low-cost, high performance CF-mMIMO systems.

In ~\cite{Zayani2023}, a local PAPR-aware precoding scheme tailored for CF-mMIMO-OFDM systems is explored. While this approach achieves notable performance enhancements, it remains limited by its reliance on a high number of antenna, which can restrict practical implementation. In \cite{Durant2024}, we introduced a team-based over the air HWI compensation solution. However, this method is developed specifically for a full-pilot zero-forcing (FZF) precoder, which is characterized by suboptimal performance in low input back-off (IBO) scenarios and incurs relatively high computational complexity. 

Post-distortion methods are a common approach to improving PA performance, particularly for amplifiers with memory effects~\cite{wang2024low}. Analog-based techniques involve designing auxiliary paths~\cite{PA_aux} or quasi-differential architectures~\cite{PA_diff} in the RF chain, increasing hardware cost and device size, key limitations for APs in CF-mMIMO. Digital-based techniques, on the other hand, are implemented at the receiver and thus restricted to uplink communications~\cite{digital_post_dist}. These methods also face computational challenges due to high-rank matrix solving, demanding high degrees of freedom for substantial performance gains~\cite{wang2024low}.

In this paper, we propose an innovative over-the-air HWI compensation method, akin to digital post-distortion, implemented at the APs but in the downlink, leveraging the unique properties of CF-mMIMO with serial fronthaul. The use of a serial fronthaul architecture enables over-the-air cooperation in CF-mMIMO networks . Successive APs in the chain apply post-distortion to compensate for the HWIs of their predecessors, with compensation becoming effective at the users through coherent signal summation. This approach balances performance and flexibility. 

\subsection{Contribution}
Our key contributions are summarized as follows:
\begin{itemize}
\item Development of a Sequential Hardware-Aware Precoding Scheme:
We propose a novel sequential precoding technique based on a partial zero-forcing (PZF) precoder that takes into account the hardware impairments (HWI) present in low-cost RF chains. Our approach effectively mitigates distortion caused by non-linear PAs, leading to enhanced spectral efficiency.

\item Development of a closed-form solution and an algorithm:
Starting from an optimization problem, we propose a closed-form solution based on a scalable sequential precoding scheme. Then, we introduce an algorithm and provide the derivation of the lower-bound spectral efficiency to evaluate the system performance using the proposed solution.

\item  Quantitative comparison:
We conduct a detailed comparison between our solution and PAPR reduction techniques taken from centralized massive MIMO literature (TR) or tailored for CF-mMIMO networks (PAPR-aware precoding). Our results demonstrate that our proposed method outperforms traditional PAPR reduction methods in terms of spectral efficiency.

\item Scalability assessment:
We provide an in-depth study of the computational complexity associated with 
the method introduced as well as for two precoding scheme, FZF and PZF. We show that our proposed solution achieves better complexity compared to TR and PAPR-aware precoding for all tested configurations. Thus, our proposed method is well-suited for practical implementation of serial fronthaul CF-mMIMO networks (e.g: radio stripes).
\end{itemize}

\section{System model}
We examine the downlink (DL) transmission in a CF-mMIMO-OFDM system, consisting of $L$ APs equipped with $M$ antennas and $K$ single-antenna users (UEs). The APs are connected to a CPU through error-free and high-capacity fronthaul links. The locations of APs and UEs are randomly and uniformly distributed within the coverage area. The standard time division duplex (TDD) protocol is adopted, where $\tau_c$ is the length of each coherence block.  
\subsection{Downlink OFDM with hardware impairments}
In the considered system, we assume that all APs serve all UEs simultaneously over the same time-frequency resources. In an OFDM-based CF-mMIMO system, following the 5G NR standard, we consider a radio frame where the time-frequency resource is divided into $N_{rb}$ resource blocks (RBs), each representing a basic unit of bandwidth and time used for data transmission. Each RB consists of $N_s$ OFDM symbols transmitted over a set of $N_{sc}=\frac{N}{N_{rb}}$ subcarriers, where $N$ is the total number of OFDM subcarriers spaced by $\Delta_f$. Let $\tau_c=N_s\times N_{sc}=\tau_p+N_u+N_d$ be the total resource units (RUs) within a RB where $\tau_p$, $N_u$ and $N_d$ correspond to the number of RUs allocated for the training, uplink (UL), and DL phases respectively. For our OFDM system, we consider a guard band of size $N_{GB}$ at both edges of the spectrum where no power is allocated for the data transmit. The sub-carriers are thus divided in two subsets:  \textit{(i)} $\Xi$, used for data transmission and \textit{(ii)} its complementary $\Xi^c$, used for guard-band.

Each coherence block experiences flat-fading and the random channel responses are statistically identical to those in any other coherence block. Additionally, the channel realizations are independent between different coherence blocks. As a result, performance analysis can be conducted by examining a single statistically representative coherence block, providing insights that apply across all blocks. We denote by $\mathbf{h}_{k,l}$ the channel between the $k$-th user and the $l$-AP on subcarrier $n$ described as 
\begin{equation}
    h_{k,l,n} = \sqrt{\beta_{k,l}} g_{k,l,n},
\end{equation}
where $n= 1, \dots, N$, $\beta_{k,l}$ and $g_{k,l,n}$ are the large-scale and small-scale fading components, respectively. 
As the frequency-domain channel coefficients remain constant for all RUs within a given RB. Then, channel estimation is performed per-RB. Moreover, the channel estimation is carried out in a distributed way at each AP using $\tau_p$ mutually orthogonal $\tau_p$-length pilot sequences emitted by users. 

The $k$-th UE sends a pilot signal $\phi_{i_k}\in \mathbb{C}^{\tau_p\times 1}$ where  $i_k\in \lbrace 1,\dots, \tau_p\rbrace$ represents the $k$-th user's pilot index. In systems with a large number of users (, i.e. $K>\tau_p$), pilot-reuse is inevitable. Thus, we denote by $\mathcal{P}_k$ the set of users who share the same pilot with the $k$-th user. The available pilots are designed to be orthogonal and normalized, i.e. $\phi_k^H\phi_{k^\prime}=0$ if $k^\prime\notin \mathcal{P}_k$ and $\phi_k^H\phi_{k^\prime}=\tau_p$ otherwise. Therefore, full-rank matrix of the frequency-domain channel estimates is given by 
\begin{equation}
    \mathbf{\bar{H}}_{l,n} = \mathbf{Y}_{l,n}\mathbf{\Phi} \in \mathbb{C}^{M\times \tau_p},
\end{equation}
where $\mathbf{Y}_{l,n}$ is the frequency-domain received pilots at the $l$-th AP on the $n$-th subcarrier and $\mathbf{\Phi}=[\mathbf{\phi}_1, \dots, \mathbf{\phi}_{\tau_p}]\in \mathbb{C}^{\tau_p\times \tau_p}$ is the used pilot book. Each AP$_l$, $\forall l$, adopt the \textit{minimum mean square error} (MMSE) channel estimation method to estimate its frequency-domain channel towards the $k$-th UE, $\forall k$, on the $n$-th subcarrier as follows
\begin{equation}
\mathbf{\hat{h}}_{l,k,n} = c_{l,k}\mathbf{\bar{H}}_{l,n}\mathbf{e_{i_k}},
\end{equation}
where $c_{l,k} \triangleq  \frac{\sqrt{\eta^u_k}\beta_{l,k}}{\tau_p\sum_{t\in\mathcal{P}_k}\eta^u_t\beta_{l,t} +1}$.
Let $\eta^u_k$ be UL normalized transmit power of the $k$-th user. Moreover, the mean-square of $\mathbf{\hat{h}}_{n,l,k}$ is given by
\begin{equation}
\gamma_{l,k} \triangleq  \frac{{\eta^u_k}\tau_p\beta^2_{l,k}}{\tau_p\sum_{t\in\mathcal{P}_k}\eta^u_t\beta_{l,t} +1}.
\end{equation}
Note that channel estimates will be used by APs to design the local precoders for DL transmission.
\subsubsection{Downlink data transmission} In the DL transmission phase, the OFDM signal transmitted on a given subcarrier is the sum of the precoded signals for each user. The overall signal transmitted by the $l$-th AP on the $n$-th subcarrier is given by
\begin{equation}\label{eq:DL_prec_signal}
    \mathbf{x}_{l,n}=\sum_{k=1}^K \sqrt{{\eta^d_{l,k}}}\mathbf{w}_{l,k,n}{\rm s}_{k,n} \in \mathbb{C}^{M\times 1},
\end{equation}
where ${\rm s}_{k,n}$ is the DL signal of the $k$-th UE on the $n$-th subcarrier ($\mathbb{E}\lbrace\lvert {\rm s}_{k,n}\rvert^2\rbrace=1$), $\mathbf{w}_{l,k,n}$ denotes the precoding vector used by the $l$-th AP and ${\eta^d_{l,k}}$ is the normalized allocated power by the $l$-th AP to the $k$-th UE. Under a total power constraint at each AP, denoted as $\eta_l^{max}$, we adopt a distributed allocation scheme at each AP based on which the power assigned to the $k$-th user at the $l$-th AP is given by (\ref{eta_l_k}).
\begin{equation}\label{eta_l_k}
\eta^d_{l,k} = \frac{\gamma_{l,k}}{\sum_{i=1}^{K}\gamma_{l,i}}\eta_l^{max}, ~\forall l, \forall k,
\end{equation}
Using this power allocation scheme, users with better channel have more allocated power.
\subsubsection{Hardware impairment model} 
Considering the large number of involved radio frequency (RF) chains in the network, CF-mMIMO systems should rely on low-cost, low-power hardware components, while retaining superior performance. HWIs could be introduced from various components. Particularly, non-linear power amplifiers (NL-PAs) are considered as significant sources of impairments. The distortions introduced by a NL-PA can be characterized by the amplitude-to-amplitude (AM/AM) and amplitude-to-phase (AM/PM) conversions \cite{Nokia}. The instantaneous input-output relationship in the NL-PA is given by (\ref{eq:NL-PA}) and is independent for each antennas in the system, and time dependent.
 \begin{equation}\label{eq:NL-PA}
    \tilde{x}_{l,m}(t)=f(x_{l,m}(t))=AM(\alpha \rho)\exp^{j(\psi+PM(\alpha\rho))}\mathbb{C}^{1\times 1} 
 \end{equation}
where $AM(.)$ and $PM(.)$ are the AM/AM and AM/PM conversions, respectively, $x_{l,m}(t)=\rho e^{i\psi}$ and $\alpha$ is a factor that scales the amplifier input power to the desired  input back-off (IBO). We have $\alpha=\frac{A_{sat}}{G \sqrt{p_{in}}}10^{\frac{-IBO[dB]}{10}}$, where $p_{in}$ is the signal average power, $G$ is the small signal gain and $A_{sat}$ is the saturation level (at the output).
According to the Bussgang's theorem, a non-linear function such as (\ref{eq:NL-PA}) can be expressed as a unique decomposition (\ref{eq:bussgang_coef}) for a given Gaussian distribution,
\begin{equation}\label{eq:bussgang_coef}
    \tilde{x}_{l,m} = \kappa_0 x_{l,m} + d_{l,m},
\end{equation}
where $\kappa$ is a linear gain of the nonlinearity, and $d$ is a zero-mean (non-Gaussian) distortion signal such that $\mathbf{x}_{l,n}$ and $d$ are uncorrelated. This theorem is true in this case since, while not valid after propagation, the distortion becomes Gaussian after OFDM demodulation among all the $N$ sub-carriers. 
In this work, we assume a perfect linearity of the PA response before saturation (as if an ideal digital pre-distortion is used) which is translated by the use of the "limiter PA" model, where the signal is clipped above $A_{sat}$ as given in (\ref{eq:clipped_signal}).

\begin{equation}\label{eq:clipped_signal}
    \tilde{a}_{l,m}(t) = 
    \begin{cases} 
    a_{l,m}(t), & |a_{l,m}(t)| \leq \frac{A_{\text{sat}}}{\mathrm{G}}, \\
    \frac{A_{\text{sat}}}{\mathrm{G}}  e^{\phi(a_{l,m}(t))}, & |a_{l,m}(t)| > \frac{A_{\text{sat}}}{\mathrm{G}},
    \end{cases}
\end{equation}

where $\mathbf{a}_{l,m}(t)$ is the time-domain signal after an IFFT is performed on $\mathbf{x}_{l,n}$, $|a_{l,m}(t)|$ is the instantaneous amplitude of the complex input signal and \( \phi(\cdot)\) represents the phase of the signal. Consequently, $d_{l,m}$ in (\ref{eq:bussgang_coef}) is given by  $d_{l,m} = \tilde{a}_{l,m}(t) - a_{l,m}(t)$ and $K=1$ in this case. The frequency domain distortion $\mathbf{d}_{l,n}$ required to compute the spectral efficiency is given by $\mathbf{d}_{l,n} = \mathcal{F}\{d_{l}\}, \quad \mathbb{C}^{M\times 1} \ $ where \( \mathcal{F}\{ \cdot \} \) denotes the discrete Fourier transform (DFT).
\subsection{Performance of Cell-Free mMIMO systems against precoding strategy}
\subsubsection{Precoding schemes}
Each AP applies a precoding scheme to its transmitted signals with the aim to maximize the spectral efficiency and minimize the interference between users. The most used precoding schemes are Maximum Ratio (MR), Full-Pilot Zero-Forcing (FZF) and Partial-FZF (PZF).
\begin{itemize}
    \item First, MR is the simplest precoding scheme aiming to maximize the signal-to-noise ratio (SNR). The precoding vector for the $k$-th user is given by 
    \begin{equation}
        \mathbf{w}^{\rm MR}_{l,k,n} = \frac{\hat{\mathbf{h}}_{l,k,n}}{\sqrt{E\lbrace \lVert \hat{\mathbf{h}}_{l,k,n} \rVert^2\rbrace}}.
    \end{equation}
    \item Full-pilot Zero-Forcing (FZF) is a more advanced precoding scheme, designed to completely eliminate interference between users. The FZF precoding vector at the $l$-th AP for the $k$-th user is obtained by taking the pseudo-inverse of the channel matrix as follows
    \begin{equation}
        \mathbf{w}^{\rm FZF}_{l,k,n} = \frac{\overline{\mathbf{H}}_{l,n}\left(\overline{\mathbf{H}}^H_{l,n}\overline{\mathbf{H}}_{l,n}\right)^{-1}\mathbf{e}_{i_k}}{\sqrt{E\lbrace \lVert \overline{\mathbf{H}}_{l,n}\left(\overline{\mathbf{H}}^H_{l,n}\overline{\mathbf{H}}_{l,n}\right)^{-1}\mathbf{e}_{i_k} \rVert^2\rbrace}},
    \end{equation}
    where $\mathbf{e}_{i_k}$ denotes the $i_k$-th column of $\mathbf{I}_{\tau_p}$. It is worth noting that the FZF scheme is limited by pilot contamination, which prevents interference cancellation between users sharing the same pilot sequence. The condition $M>\tau_p$ is crucial to the success of FZF in CF-mMIMO systems, as it ensures that each AP has sufficient degrees of freedom to manage interference.
    \item PZF precoding scheme combines the last two techniques by applying partial interference suppression for the most impactful users, while maximizing the transmitted signal strength for other users via MR precoding scheme. The PZF precoding vector at the $l$-th AP for the $k$-th user is formulated as follows
    \begin{align} \label{eq:PZF_prec}
    \vspace{-0.1cm}
        &\mathbf{w}^{\rm PZF}_{l,k,n}=\\
        \nonumber
        &\begin{cases}
            \mathbf{w}^{\rm PZF}_{l,k,n}=\frac{\mathbf{\overline{H}}_{l,n}\mathbf{E}_{\mathcal{S}_l}\left(\mathbf{E}^H_{\mathcal{S}_l}\mathbf{\overline{H}}^H_{l,n}\mathbf{\overline{H}}_{l,n}\mathbf{E}_{\mathcal{S}_l}\right)^{-1}\varepsilon_{j_{l,k}}}{\sqrt{\frac{1}{\left(M-\tau_{\mathcal{S}_l}\right)\theta_{l,k}}}}, & k \in \mathcal{S}_l \\
            \mathbf{w}^{\rm MR}_{l,k,n}=\frac{\mathbf{\overline{H}}^H_{l,n}\mathbf{e}_{i_k}}{\sqrt{M\theta_{l,k}}}, & k \in \mathcal{W}_l \\
        \end{cases},
    \end{align}
    where $\mathcal{S}_l$ is the set of UEs with the best channel conditions to the $l$-th AP and considered as strong while the remaining UEs construct the set of weak UEs denoted by $\mathcal{W}_l$. Users are grouped based on their path-loss as follows : Each AP begins by finding the number of strong users that contribute a certain percentage, $\nu_{l,th}\%$, of the total channel gains from all users active in the coverage area.
    \begin{equation}
        \label{grouping}\sum_{k=1}^{\tau_{\mathcal{S}_l}}\frac{\overline{\beta}_{l,k}}{\sum_{t=1}^K \beta_{l,t}}\geq \nu_{l,th}\%,
    \end{equation}
\end{itemize}
where $\tau_{\mathcal{S}_l}$ denotes the number of users considered as strong by the $l$-th AP and $\lbrace \overline{\beta}_{l,1},\dots, \overline{\beta}_{l,K}\rbrace$ is the set of large-scale fading coefficients sorted in the descending order. Hence, users sharing the same pilot with a strong user are also considered as strong users. We also have $ \tau_{\mathcal{S}_l}\leq \min(M,\tau_p+1)$.

\subsubsection{Performance of the system}
By tuning grouping threshold in (\ref{grouping}), we can emulate the behavior of the FZF precoder ($\nu_{l,th} = 100\%$). Therefore, we only derive the performance for the PZF precoder in this section but we give results equivalent to those of the FZF precoder in section \ref{sec:Numerical_Results}. The  frequency-domain received signal at UE $k$ at sub-carrier $n$ can be given by (\ref{y_k_n_exp}).
\begin{equation}\label{y_k_n_exp}
\begin{aligned}
y_{k,n} &= \underbrace{\sum_{l=1}^{L}\sqrt{\eta^d_{l,k}}\mathbf{h}^H_{l,k,n}\mathbf\mathbf{w}_{l,k,n}s_{k,n}}_{\text{Desired signal}} + \underbrace{b_{k,n}}_{\text{Noise}} + \\
& \underbrace{\sum_{l=1}^{L}\sum_{t\neq k}^{K}\sqrt{\eta^d_{l,t}}\mathbf{h}^H_{l,k,n}\mathbf\mathbf{w}_{l,t,n}s_{t,n}}_{\text{Multi-user interference}} + \underbrace{\sum_{l=1}^{L}\mathbf{h}^H_{l,k,n}\mathbf{d}_{l,n}}_{\text{HWI}} 
\end{aligned}
\end{equation}
where $\mathbf{d}_{l,n} \in \mathbb{C}^{M\times 1}$ is the frequency-domain version of hardware-related distortion and $b_{k,n} \sim \mathcal{CN}(0,1)$ is an i.i.d. Gaussian noise.

The per-user SE in the downlink can be computed \cite{Interdonato2020} as 
\begin{equation}\label{SE_k1}
\text{SE}_{k} = \xi\left(1 - \frac{\tau_p}{\tau_c}\right)\text{log}_2\left(1 +  \text{SINR}_{k,n}\right),  
\end{equation}
with the effective signal-to-interference-plus-noise ratio (SINR)  of UE $k$ at sub-carrier $n$, with a normalization by the noise power, is given by
\begingroup\small
\begin{align}\label{SINR_kn}
\centering
&\text{SINR}_{k,n}  \notag \\
&= \frac{\left|\mathrm{CP_{k,n}}\right|^2}{\mathrm{E}\left\{\left|\mathrm{PU_{k,n}}\right|^2\right\}+\sum_{t\neq k}^{K}\mathrm{E}\left\{\left|\mathrm{UI_{k,t,n}}\right|^2\right\}+\mathrm{E}\left\{\left|\mathrm{HWI_{k,n}}\right|^2\right\}+1}
\end{align}
\endgroup
where $\mathrm{CP_{k,n}}$, $\mathrm{PU_{k,n}}$, $\mathrm{UI_{k,n}}$ and $\mathrm{HWI_{k,n}}$ are the coherent precoding, the precoding gain uncertainty, the multi-user interference and the hardware impairment respectively, and are given by (\ref{eq:CP}, \ref{eq:PU}, \ref{eq:UI}, \ref{eq:HWI}).

\begin{equation}\label{eq:CP}
    \mathrm{CP_{k,n}} = \sum_{l=1}^{L}\sqrt{\eta^d_{l,k}}\mathrm{E}\left\{\mathbf{h}^H_{l,k,n}\mathbf\mathbf{w}_{l,k,n}\right\}
\end{equation}
{\small
\begin{equation}\label{eq:PU}
   \mathrm{PU_{k,n}} = \sum_{l=1}^{L}\left(\sqrt{\eta^d_{l,k}}\mathbf{h}^H_{l,k,n}\mathbf\mathbf{w}_{l,k,n}-\sqrt{\eta^d_{l,k}}\mathrm{E}\left\{\mathbf{h}^H_{l,k,n}\mathbf\mathbf{w}_{l,k,n}\right\}\right)  
\end{equation} }
\begin{equation}\label{eq:UI}
    \mathrm{UI_{k,t,n}} = \sum_{l=1}^{L}\sqrt{\eta^d_{l,k}}\mathbf{h}^H_{l,k,n}\mathbf\mathbf{w}_{l,t,n}
\end{equation} 
\begin{equation}\label{eq:HWI}
    \mathrm{HWI_{k,n}} = \sum_{l=1}^{L}\mathbf{h}^H_{l,k,n}\mathbf{d}_{l,n}
\end{equation}
with $i_k$ and $i_t$ the indices of the pilot sequence used by UE $k$ and $t$. It corresponds to the column vector in the precoder matrix designed for their training sequences.
\section{Compensation of HWI : PAPR reduction}

\subsection{Tone reservation}\label{AA}
Tone Reservation (TR) is a widely utilized technique for reducing the PAPR in OFDM systems.In TR, a subset of subcarriers \( \mathcal{R} \subseteq \{N_{GB}, 1, \dots, N_{sc}-N_{GB}\} \) with size $N_{TR}$, known as reserved tones, is allocated for PAPR reduction. These reserved tones carry no data and are optimized iteratively by injecting specially designed signals, such as clipping noise, to cancel out high peaks in the time-domain signal. These signals are constructed to minimize interference with the data-carrying subcarriers, thereby preserving the integrity of the transmitted information. One notable advantage of the TR method is that the reserved tones can be set to zero at the receiver without affecting data demodulation.
Let \( \mathbf{c}_{l,m} \) represent the clipping noise introduced on the reserved tones. The frequency domain signal is given by (\ref{eq:TR_freq}),
\begin{equation}\label{eq:TR_freq}
\overline{\mathbf{x^i}}_{l,m} = \mathbf{x}^{\text{data}}_{l,m} + \mathbf{c}_{l,m},    \in \mathbb{C}^{N_{sc}\times 1}
\end{equation}
where indices $i$ represent the $i$-th iteration and \( \mathbf{x}^{\text{data}}_{l,m} \) contains the frequency domain precoded DL signal on non-reserved subcarriers such that \( \mathbf{x}^{\text{data}}_{l,m} =  \mathbf{x}_{l,m} \forall n \notin \mathcal{R}\), and \( \mathbf{c}_{l,m} \) is non-zero only for \( n \in \mathcal{R} \).

Then, the precoded signal, $\mathbf{x}_{l,m}$, is converted into the time-domain signals. The resulting signal, $\mathbf{\overline{a^{i}}_{l,m}}^\text{TR}$ given by (\ref{eq:x_Tr}), is used to compute the clipping noise. 
\begin{equation} \label{eq:x_Tr}
    \mathbf{\overline{a^i}}^\text{TR}_{l,m} = \mathbf{a}^{\text{data}}_{l,m} + \frac{1}{\sqrt{N_{sc}}} \sum_{n \in \mathcal{R}} \mathbf{c}_{l,m}(n) e^{j2\pi n/N_{sc}}
\end{equation}
The objective is to determine \( c_{l,m} \) such that the peaks of \( |\mathbf{x^\text{TR}}_{l,m}| \) are minimized. The clipping noise \( \mathbf{C} \) is updated based on the instantaneous time-domain signal  (at time $\mathit{t}$, with number of sample set by the IFFT) exceeding a predefined threshold \( T \). The clipping noise is computed as:
\begin{equation}\label{eq:clipping}
\mathbf{c}_{l,m} = 
\begin{cases} 
\mathbf{a}_{l,m}(t) - T e^{j\phi(\mathbf{a}_{l,m}(t))}, & \text{if } |\mathbf{a}_{l,m}(t) | > T, \\
0, & \text{otherwise}
\end{cases}
\end{equation}
The clipping noise is then projected back onto the reserved tones $n$ in the frequency domain as
\begin{equation}\label{eq:clipping_freq}
 C_{l,m}(n) = 
\begin{cases}
\mathcal{F}\{c_{l,m}(t)\}, & n \in \mathcal{R}, \\
0, & n \notin \mathcal{R},
\end{cases}   
\end{equation}
The process is then repeated  from (\ref{eq:x_Tr}), with the previous output $\mathbf{\overline{a}^{(i-1)}}^\text{TR}_{l,m}$ replacing $\mathbf{a}^{\text{data}}_{l,m}$ for the $N^{TR}_{it}$ subsequent iterations. The effectiveness of TR depends on the choice of \( \mathcal{R} \), the threshold \( T \), and the number of iterations  $N^{TR}_{it}$.

Here, we adopt a parameter-based threshold design strategy (\ref{eq:Threshold_opti}), with the optimal value of \( T \) analytically derived in \cite{Jiang2014} as follows

\begin{equation}\label{eq:Threshold_opti}
T = P_{\text{mean}[\mathbf{x}^{\text{data}}_{l,m}]} \cdot \sqrt{\ln\left(\frac{N_{\text{sc}}}{|\chi|}\right)}
\end{equation}
with $P_{\text{mean}[\mathbf{x}^{\text{data}}_{l,m}]}$ the mean power (on active sub-carrier) of $\mathbf{x}^{\text{data}}_{l,m}$ and $\chi$ is the number of sub-carrier used to store the distortion, here, $\chi = N_{TR}$ .
\subsection{Local PAPR-aware  Precoding}
To mitigate the effects of NL-PAs, we consider the clip control (CC) approach for PAPR reduction. By implementing a PAPR-aware local precoding, the method ensures that the power peaks of transmitted signals are controlled locally, minimizing the non-linear distortions caused by APs. The key idea of the proposed approach is to send low-PAPR signals  by generating the Peak Canceling Signals (PCSs) $\{\mathbf{\mathsf{r}}_{l,n}, \forall l \forall n\}$. These signals are adaptively generated and added to transmitted signals from each AP to suppress excessive power peaks while minimizing the impact on overall signal quality. Thus, the effective frequency-domain version of transmit signal vector on the $n$-th subcarrier at the $l$-th AP is given by
\begin{equation}
    \overline{\mathbf{x}}_{l,n}=\mathbf{x}_{l,n}+\mathbf{r}_{l,n},
\end{equation} 
where $\mathbf{r}_{l,n}\in \mathbb{C}^{M\times 1}$ is the frequency-domain PCS vector at the $n$-th subcarrier. It is worth mentionning that optimal PCSs should correspond to the clipping noise in the frequency domain $C_{l,m}(n)$, computed as in the TR method with (\ref{eq:clipping}) and (\ref{eq:clipping_freq})


Frequency-domain PCSs are optimized such that distortion caused to the strongest users is negligible or even null while distortion caused to the weakest users are tolerable.Therefore, we have 
\begin{equation}
    \mathbf{r}_{l,n} = \omega_l\mathbf{V}_{l,n}\mathbf{q}_{l,n},
\end{equation}
where $\mathbf{V}_{l,n}$ is the projection matrix used at the $l$-th AP on the $n$-th subcarrier to cancel distortions towards the strong UEs. Additionally, $\omega_l$ is a regularization factor. To generate PCSs that are close to the original clipping noise in \eqref{eq:clipping_freq},  it is necessary to simultaneously optimize $\omega_l$ and $\mathbf{q}_{l,n}$. Hence, the optimization objective can be formulated as follows
\begin{equation}
\begin{aligned}
\underset{(\omega_l,\mathbf{q}_{l,n})}{\mathrm{arg\,min}} \quad & \lVert \omega_l\mathbf{V}_{l,n}\mathbf{q}_{l,n}-\mathbf{C}_{l,n} \rVert^2_2 \\
\textrm{s.t.} \quad & {\rm s}_{k,n}=\mathbf{h}^H_{l,k,n}\left(\mathbf{x}_{l,n}+\omega_l\mathbf{V}_{l,n}\mathbf{q}_{l,n}\right),~\forall k\in\mathcal{S}^{\rm PAPR}_l (C1)\\
& \mathbb{E}\left[ \|\overline{\mathbf{x}}_{l,n}\|^2_2 \right] \leq \eta^{\text{max}}_{l,n} \quad \forall l,n  (C2)\\
  & \mathbf{r}_{l,n}=\mathbf{0}_{M\times1},~\forall n\in\Xi^c (C3)\\
\end{aligned}
\end{equation}
The constraints (C1), (C2) and (C3) play a key role in ensuring that PCSs meet system requirements.  The first constraint (C1) ensures that PCSs are null towards strong users ($\mathcal{S}^{\rm PAPR}_l$). The second constraint (C2) ties the precoder design to the maximum per-AP power constraint. Finally, (C3) is designed to prevent the out-of-band (OOB) radiation. It ensures that PCSs do not produce any energy in the OOB subcarriers defined by $\Xi^c$.

We propose an iterative process is designed to make the PCSs fit to their associated frequency-domain clipping-noise signals.
\begin{table}[h!]
\begin{tabular}{l}
\hline
\textbf{Algorithm I}: The Local PAPR-aware Precoding algorithm\\
\hline
Input:\\
- a set of $N$ modulated complex signals ~$\{\mathbf{s}_{k}\}$\\
- the channel estimates $\{\mathbf{\bar{H}}_{l,n}\}, \forall~n=1,..,N, \forall~l=1,..,L$\\
- the UEs path-losses ~$\{\beta_{l,k}\}, \forall~l=1,..,L,  \forall~k=1,..,K$\\
1: \textbf{Define}:\\
- the two groups of users {(\textit{strong} and \textit{weak})} based on $\nu^{\rm PAPR}_{th}$\\
- the corresponding regularization diagonal matrix ~$\mathbf{\Gamma}_l$ \\
2: \textbf{Compute}:\\
- the projection matrices $\{\mathbf{V}_{l,n}\}$ \\
- the precoded data vectors $\{\mathbf{x}_{l,n}\}$ \\
3: \textbf{Set} the maximal iteration number $N_{iter}$\\
4: \textbf{Initialize}:\\
- $\{\mathbf{r}^{(1)}_{l,n} = \mathbf{0}_{M\times 1}\}$\\
- $\{\mathbf{x}^{(1)}_{l,n}\} = \{\mathbf{x}_{l,n}\}$\\
5: \textbf{for} ~p=1,...,$N_{iter}$ ~\textbf{do} \\
6: $\mathbf{\mathsf{a}}^{t(p)}_{m} = {\rm IFFT}\left(\mathbf{x}^{t(p)}_{m}\right), ~ \forall ~ m=1...M$ \\
7: $\mathbf{\epsilon}^{(p)}_{l,n} = {\rm FFT}\left(\mathbf{\bar{\mathsf{a}}}^{t(p)}_{m} -\mathbf{\mathsf{a}}^{t(p)}_{m} \right)$ \\
8: $\mathbf{\epsilon}^{(p)}_{l,n} = \mathbf{0}_{M\times 1}$, for $n\in\chi^c$\\
9: $\omega_l=\mathrm{E}_n\left[\frac{\sum_{m}|\mathbf{V}_{l,n}\mathbf{\epsilon}^{(p)}_{l,n}||\mathbf{\epsilon}^{(p)}_n|}{\sum_{m}{|\mathbf{V}_{l,n}\mathbf{\epsilon}^{(p)}_{l,n}|}^2}\right]$ \\
10: $\mathbf{r}^{(p)}_{l,n} = \omega_l\mathbf{V}_{l,n}\mathbf{\epsilon}^{(p)}_{l,n}$,~ $\forall$ ~$n\in\Xi$ \\
11: $\mathbf{x}^{(p+1)}_{l,n} = \mathbf{x}^{(p)}_{l,n} + \mathbf{r}^{(p)}_{l,n}, ~ \forall n\in\Xi$ \\
12: \textbf{end for}\\
13: \textbf{return} $\{\mathbf{x}^{N^{PAPR}_{it}}_{l,n}\}$\\
\hline
\end{tabular}
\end{table}
The proposed algorithm is based on the following steps
\begin{enumerate}
    \item Defining the groups of users (step 1) : The same grouping strategy used in \eqref{grouping} for the PZF precoding is used here but with a different threshold $\nu_{l,th}^{\rm PAPR}$.
    \item Projection matrix computation (step 2):
    \begin{equation}\label{eq:FPS}
    \mathbf{V}_{l,n} = \mathbf{I}_M  - \mathbf{\bar{H}}_{l,n}\left(\mathbf{\bar{H}}^H_{l,n}\mathbf{\bar{H}}_{l,n} + \mathbf{\Gamma}_{l}\right)^{-1}\mathbf{\bar{H}}^H_{l,n}.
    \end{equation}
    Note that we have $\mathbf{\omega}_l\mathbf{\hat{h}}_{l,k,n}\mathbf{V}_{l,n}\mathbf{\epsilon}_{l,n} = 0$, only if $\mathbf{\Gamma}_{l}\left[k,k\right]=0$, meaning that distortion caused to the $k$-th UE is almost removed. Thus, $\mathbf{\Gamma}_{l}\left[k,k\right] = 0, ~\forall k ~\in~ \mathcal{S}^{\rm PAPR}_l$.

    \item Initialization (step 4) : The PCSs are initialized to zero for each AP and subcarrier. The initial precoded signals are the foundation for designing the PCSs.
    \item PCSs Computation (step 10):
    \begin{itemize}
        \item (step 6 \& 7): In step 6, similarly to the TR method, the precoded signal $\mathbf{x}^{(p)}_{l,n,m} $ is converted into the time-domain signals and used in (step 7) to compute the clipping noise using \eqref{eq:clipping} and \eqref{eq:clipping_freq}.
   The optimal threshold for the clipping is the same as in \eqref{eq:Threshold_opti}, with $\Xi = N_{sc} - 2BG$.
        \item  (step 8 \& 9): Step 8 is used to remove the OOB radiations. Then, the regularization factor is computed in a way to adjust the PCSs close to the original clipping noise.
    \end{itemize}
    \item Precoded signal update (step 11): Once PCSs are designed, they are added back to the precoded signals.
\end{enumerate}
This process is repeated until reaching the maximum number of iterations $N^{PAPR}_{it}$.

\section{Sequential Hardware-Aware Precoding}
The TR and PAPR-aware precoder methods are limited in the achievable performance gain. For the TR, performance are proportional to the number of reserved tone, which reduces the available spectral range for DL, and the number of iterations of the algorithm to enhance the PAPR reduction. The PAPR-aware also requires multiple iterations on top of the necessity of having large number of antennas to design an optimal solution. Moreover, they present high computational complexity and are more limited in providing benefits for the non-linearity outside of the saturation region and requires an additional linearization such as DPD or highly linear PAs.

\begin{figure}[ht!] 
    \centering
    \includegraphics[width=1\linewidth]{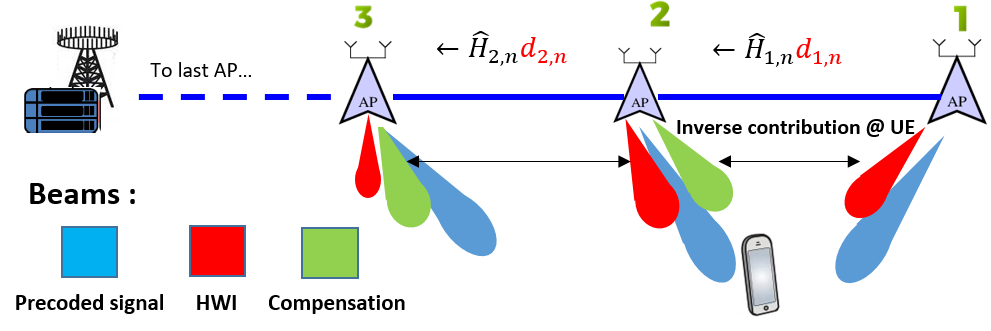}
    \caption{Graphical representation of several APs connected through a serial fronthaul link. With the sequential HW-aware precoder, the distortions of any AP $l-1$ are canceled at the location of the user in an over-the-air manner. }
    \label{figure:Sequenital_schematic}
\end{figure}

In this context, we propose to tackle the problematic of PA impairment in a different way. With respect to the constraints of local precoding, thanks to the collaborative aspect of serial fronthaul topologies in CF-mMIMO, we can design sequential over-the air HWI cancellation solution. We illustrate the sequential compensation process on Figure \ref{figure:Sequenital_schematic}.  In our solution, each AP independently estimates its local channel during the uplink training phase and approximates the distortions \(\{\mathbf{d}_{l,n}, \forall n\}\) using the transfer function model of the hardware (\ref{eq:bussgang_coef}). The precoding process begins at the first AP (\(l = 1\)), which generates the transmitted signal using a the precoder \(\mathbf{W}^{\rm PZF}_{l,n}\) defined in (\ref{eq:PZF_prec}). Once the signal is generated, the first AP approximates and forwards the user-perceived HWI \(\{\mathbf{\hat{H}}^H_{1,n}\mathbf{d}_{1,n}, \forall n\}\) to the next AP along the serial fronthaul. Each subsequent AP (\(l > 1\)) receives distortion information from the previous AP (\(l-1\)) and combines it with its own channel estimates \(\{\mathbf{\hat{H}}^H_{l,n}, \forall n\}\) to refine the precoding strategy for its own transmission.   These latter will enable, at the $K$ UEs antennas, a coherent over-the air construction of useful signals and also the over-the air cancellation of the hardware-related distortions caused by the preceding AP.

The precoding optimum for any AP $l$ when the distortions are included can be written as an optimization problem (\ref{eq:Condition_design_HWaw}),
\begin{equation}\label{eq:Condition_design_HWaw}
\begin{aligned}
\underset{\boldsymbol{\Omega}_{l,n}\mathbf{W}_{l,n}}{\mathrm{arg\,min}} \quad & \left\lVert \mathbf{H}_{l,n}\mathbf{W}_{l,n}\mathbf{s}_{k,n} + \mathbf{\Omega}_{l,n} \mathbf{H}_{l-1,n}\mathbf{d}_{l-1,n} - \mathbf{s}_{k,n} \right\rVert^2_2 \\
\textrm{s.t.} \quad & \mathbb{E}\left[ \| \mathbf{x'}_{l,n} \|^2_2 \right] \leq \eta^{\text{max}}_{l,n}, \quad \forall l,n \quad \text{(C1)} \\
& \mathbf{x'}_{l,n}=\mathbf{0}_{M\times1},~\forall n\in\Xi^c \quad \text{(C2)}\\
\end{aligned}
\end{equation}
where $\mathbf{x}^\prime_{l,n}$  is the the final transmitted frequency-domain precoded signal.
$(C1)$  constrains the maximum power output with per-AP power constraint for the final transmitted signal and $(C2)$ limits the out of band radiations.

 In order to solve \eqref{eq:Condition_design_HWaw}, the proposed precoding scheme consists of two main stages. First, the primary precoder \(\mathbf{W}^{\rm PZF}_{l,n}\) is computed to perform spatial multiplexing. Then, a secondary precoder \(\mathbf{W}'_{l,n} \in \mathbb{C}^{M \times K}\) is designed to mitigate the HWI. The final transmitted frequency-domain precoded signal \(\mathbf{x}^\prime_{l,n}\) from AP \(l\) to the \(K\) UEs on subcarrier \(n\) is given by \eqref{eq:x_l_n_serial}.

\begin{equation}\label{eq:x_l_n_serial}
\mathbf{x}^\prime_{l,n} =
\begin{cases} 
 \mathbf{W}^{\rm PZF}_{l,n}\mathbf{P}_l\mathbf{s}_n  &  l=1
\\
 \mathbf{W}^{\rm PZF}_{l,n}\mathbf{P}_l\mathbf{s}_n - \mathbf{W'}_{l,n}\mathbf{\hat{H}}^H_{l-1,n}\mathbf{d}_{l-1,n} & \forall l>1
\end{cases}
\end{equation}
where $\mathbf{P}_l$ is the diagonal power allocation matrix containing the $K$ elements $\eta^d_{l,k}$ defined in \eqref{eta_l_k}.

In order to meet the power constraint requirement (C1) in \eqref{eq:Condition_design_HWaw} we design the secondary precoder matrix \(\mathbf{W'}_{l,n}\) as in (\ref{eq:W_prime}) following the PZF approach paired with a regularization term $\sqrt{\frac{1}{\beta_{l,k}}}$ for power scaling. 

\begin{equation}\label{eq:W_prime}
w^{'}_{l,n,k} =
\begin{cases} 
 \sqrt{\frac{\theta_{l,k}}{\beta_{l,k}}}\mathbf{\hat{H}}_{l,n} \left(\mathbf{\hat{H}}^H_{l,n} \mathbf{\hat{H}}_{l,n}\right)^{-1} \mathbf{e}_{i_k}  & \forall k \in S_{l}, \\
0 & \forall k \in W_{l},
\end{cases}
\end{equation}
The HW-aware precoding with PZF \eqref{eq:W_prime} is applied only to users that exhibit favorable channel conditions, i.e the strong users. The objective of the technic is to compensate the HWI, to protect the strong user, and continue to offer service to weak users. The frequency-domain received signal at UE $k \in \mathcal{S}_l $ and subcarrier $n$ is given by (\ref{y_k_n2}). 
\begin{figure*}[h!]
\begin{equation}\label{y_k_n2}
\begin{aligned}
y_{k,n} = \sum_{l=1}^{L}\mathbf{h}^H_{l,k,n}\kappa_0\mathbf{x}_{l,n} + \sum_{l=1}^{L}\mathbf{h}^H_{l,k,n}\mathbf{d}_{l,n} + b_{k,n} \underbrace{\sum_{l=1}^{L}\sqrt{\eta_{l,k}}\mathbf{h}^H_{l,k,n}\kappa_0\mathbf{w}_{l,k,n}s_{k,n}}_{\text{Desired signal}} + \underbrace{\sum_{l=1}^{L}\sum_{t\neq k}^{K}\sqrt{\eta_{l,t}}\mathbf{h}^H_{l,k,n}\kappa_0\mathbf{w}_{l,t,n}s_{t,n}}_{\text{Multi-user interference}} \\
+ \underbrace{\sum_{l=1}^{L-1}\left[\mathbf{h}^H_{l,k,n} - \mathbf{h}^H_{l+1,k,n}\mathbf{W}_{l+1,n}\mathbf{\hat{H}}_{l,n} \right]\mathbf{d}_{l,n} + \mathbf{h}^H_{L,k,n}\mathbf{d}_{L,n}}_{\text{Residual HWI}} + \underbrace{b_{k,n}}_{\text{Noise}} \;\;\;\; \forall k \in \mathcal{S}_l
\end{aligned}
\end{equation}
 \hrulefill
\end{figure*}

The scheme is designed with minimal signaling overhead, with only \(MN\) complex samples exchanged between APs. The signaling requirements remain constant along the fronthaul, and the complexity of the secondary precoder step is proportional to the size of the set \(S_{l}\), which typically contains only a small fraction of the users.
   
\
The residual HWI, denoted as \(\mathrm{HWI'_{k,n}}\), related only to the channel estimation error, is expressed in (\ref{psi2}). This new term replace the contribution (\ref{eq:HWI}) in (\ref{SINR_kn}) in the evaluation of the system performance. Notably, it applies only to UEs in \(S_{l}\). 

\begin{align}\label{psi2}
  \mathrm{HWI'_{k,n}} &= \sum_{l=1}^{L-1}\left[\mathbf{h}^H_{l,k,n} - \mathbf{h}^H_{l+1,k,n}\mathbf{W'}_{l+1,n}\mathbf{\hat{H}}_{l,n} \right]\mathbf{d}_{l,n} \nonumber \\
&+ \mathbf{h}^H_{L,k,n}\mathbf{d}_{L,n} \; \forall k \in S_{l}
\end{align}

The term \(\mathbf{h}^H_{L,k,n}\mathbf{d}_{L,n}\) represents the uncompensated distortion introduced by the final AP $L$ in the branch. To minimize this distortion, it is essential that the last AP operates with sufficient resources and high IBO for its PA, ensuring that the distortion effects are kept to a minimum.

Through the sequential sharing of distortion information and adaptive precoding strategies, the solution effectively cancels the impact of hardware impairments while maintaining high system performance. This approach not only addresses HWI at each AP but also ensures that the computational and signaling overhead remains manageable, enabling practical deployment in large-scale CF-mMIMO systems.

\begin{table*}[ht]
\centering
\caption{Complexity Analysis of HWI Compensation Methods in Terms of Number of Complex Multiplications}
\begin{tabular}{|c|c|c|c|}
\hline
\textbf{Method} & \textbf{Key Operations} & \textbf{Complexity (per channel realization)}  \\ \hline
Partial Zero-Forcing (PZF) & 
Matrix inversion \& multiplication + application &
$O(|\Xi|\cdot(MK+M\tau_s^2+\tau_s^3+M\tau_s^2)+N_s\cdot(|\Xi|(MK))$
\\ \hline
Tone Reservation (TR) & 
FFT/IFFT over $N^{TR}_{it}$ iter. &
$O(N_s\cdot(MN^{TR}_{it}\cdot(2|\Xi|log(|\Xi|))))$ \\ \hline

PAPR-aware Precoding & 
Matrix projection, IFFT/FFT over $N^{PAPR}_{it}$ iter. &
$O(N^{PAPR}_{it}\cdot (2M|\Xi|log(|\Xi|)+|\Xi|M^2)+ (K^3 +2(K^2M)+M^2K))$ \\ \hline

Hardware-aware Precoding & 
Matrix projection, FFT/IFFT &
$O(M\cdot |\Xi|\tau_s+N_s\cdot((M\cdot\tau_s+M^2\cdot\tau_s+M^2) + 3M\cdot |\Xi|log(|\Xi|)))$  \\ \hline

\end{tabular}
\label{tab:complexity_analysis}
\end{table*}

\section{Complexity Analysis}

 First, we would like to compare the complexity of the PZF and the FZF precoder. The precoder design is performed once per coherent channel interval and its complexity is mostly dependent on the size of the matrix inversion and multiplication applied to the set $\mathcal{S}_l$ in (\ref{eq:PZF_prec}), while the MRT-like computation for the set $\mathcal{W}_l$ is negligible in terms of added complexity. If all users are strong (\(\tau_s = K\)), the complexity is equivalent to a full-pilot zero forcing (FZF). The precoder is then applied to the $N_s$ OFDM symbols of the coherence block following (\ref{eq:DL_prec_signal}). The resulting total complexity is given in Table \ref{tab:complexity_analysis}. Then, we evaluate the computational complexity of the three solutions : TR, PAPR-aware precoding, and hardware-aware precoding.  The Tone Reservation method is primarily influenced by the IFFT/FFT operations and the clipping process to minimize the PAPR. Specifically, the IFFT and FFT operations both require \( O(M \cdot\|\Xi| \log(|\Xi|)) \) complex multiplications where \( |\Xi| \) is the number of active sub-carrier. The clipping operation is performed only with addition or subtraction, thus its complexity is neglected. Most notably, the iterative nature of the operation brings $N^{TR}_{it}$ repetitions of IFFT/FFT operations, which is computationally heavy. The PAPR-aware precoding method can be decomposed in three mains steps : the computation of the projection matrix $V_{l,n}$ to cancel distortions towards strong UEs once per channel block, the computation of the clipping noise in the time domain to generate the PCS signal over $N^{PAPR}_{it}$ iterations, and finally the projection to generate the updated precoded signal. The total complexity is given in Table \ref{tab:complexity_analysis}.  The balance between complexity and performance can be tuned by selecting an adequate number of iteration for the PAPR-aware and Tone Reservation methods, as it affect their computational load. On the other hand, the sequential HW-aware method is not locally iterative and always requires three steps to operate :
\begin{enumerate}
    \item To calculate the new precoder weigth in $\mathbf{W'}_{l,n}$ (once per channel)
    \item To perform the projection \(\mathbf{W'}_{l,n}\mathbf{\hat{H}}^H_{l-1,n}\mathbf{d}_{l-1,n}\) and subtract it to the precoded signal $\mathbf{W}^{\rm PZF}_{l,n}\mathbf{P}_l\mathbf{s}_n $ (once per symbol)
    \item To evaluate the local HWI using the transfer function of the PA (\ref{eq:clipped_signal}) (once per symbol)
\end{enumerate}
  Obtaining the complexity of 1) is straight-forward as the matrix inversion and multiplication performed in the numerator of (\ref{eq:PZF_prec}) can be re-used, thus, only the normalization term (scalar) is redesigned. It results in \(M\cdot |\Xi|\tau_s\) additional complex multiplications to apply the scaling $\sqrt{\theta_{l,k}}$ . For 2), the projection requires \((M\cdot\tau_s+M^2\cdot\tau_s+M^2)\) operations to be performed, and the subtraction is neglected due to its faster nature. Finally, the computation of the HWI is constrained mainly by two IFFT/FFT costing \((2M\cdot |\Xi|log(|\Xi|))\) multiplications, while \((2M\cdot |\Xi|)\) are required to compute the transfer function of the PA (\ref{eq:clipped_signal}). Remarkably, the total complexity of this method depends of the complexity of the PA model considered to translate the impact of the real component. A balance must be found between accurate modeling and the computational complexity of the model.  It is also worth noticing that this method use the information feed link to share the local CSI and HWI estimates to the next AP in the chain. It doesn't put more efforts on the local CPU but requires the transmission of a $|\Xi|\mathrm{x} M \mathrm{x} K$ sized and a $ |\Xi| \mathrm{x} K$ sized complex matrix along the fronthaul.

\section{Simulation : parameters \& scenario}

\subsection{Environment}

In this work, we consider a square simulation environment measuring 500 x 500 meters, representing a factory-like type of indoor scenario. It contains $L=200$ APs uniformly distributed along the central area of the environment at equal intervals, positioned on the roof at a height of 10 meters. Meanwhile, $K$ UEs are randomly positioned at a height of 1.5 meters. This arrangement ensures consistent coverage across the area. 
\subsection{System parameters}
The simulation parameters are outlined in Table \ref{tab:SimSet}. The maximum per-AP downlink transmit power, $\eta^{max}$, is constrained by the IBO to standardize the performance across comparable HWI levels, calculated as $\eta^{max} =\mathrm{M} (\mathrm{A_{sat}}/\mathrm{G})10^{\mathrm{IBO}/10}$. A total of 400 snapshots are simulated, with 25 random sequences (including M-QAM signals and channel realizations) generated for each snapshot.

\begin{table}[bh]
    \begin{center}
    \begin{tabular}{ l | l || l | l }
    \textbf{Description} & \textbf{Value} & \textbf{Description} & \textbf{Value} \\
    \hline
    Noise power & $-93$ dBm & UL transmit power & 20 dBm \\ 
    $\mathrm{A_{sat}}$ & 1.9 V & G & $16$\\ 
    $\tau_c$  & $168$ & $\xi$ & $0.5$  \\
    $N_{chan}$ & 25 & $\sigma_{sh}$ & $4$ dB \\
    $f_c$ & $3.5$ GHz & Bandwidth & $20$ MHz \\
    N & 256 & M & 8\\
    \end{tabular}
    \caption{\label{tab:SimSet}Simulation settings.}
    \end{center}
\end{table}

\section{Numerical results}\label{sec:Numerical_Results}

\subsection{Cell-Free massive MIMO under hardware impairments}

\begin{figure}[ht!] 
    \centering
    \includegraphics[width=0.85\linewidth]{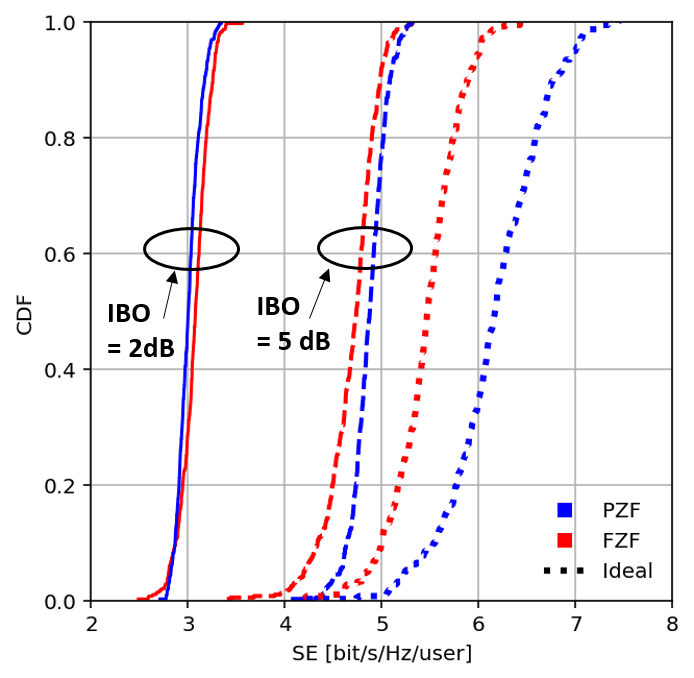}
    \caption{CDF of the SE for the FZF and the PZF precoder ($\nu_{th}=99\%$) for two values of IBO : IBO$ = 2$dB and IBO $= 5$ dB. Ideal performance between different IBO values are equal due to the normalization of the transmit power w.r.t the IBO, here $\eta^{DL}_{l,k}= 18.52 \mathrm{dBm}$}
    \label{fig:CDF_1}
\end{figure}

First, the impact of the “limiter-PA” on the performance of the CF-mMIMO systems is studied. In Figure \ref{fig:CDF_1}, the CDF of the spectral efficiency are given for different values of IBO: 2 dB for a high level and 5 dB for a moderate level of distortions. It must be noted that a reference scenario with $K=\tau_p=7$, $M=8$ and $L = 200$ AP (a distance of 25 m between two APs) is used if not said otherwise. From the results, it is clear that the HWI are dominating the negative contributions in the SINR (\ref{SINR_kn}). With the 2 dB IBO scenario, only up to 3.4 bit/s/Hz/user are obtained as opposed to the maximum achievable SE with ideal hardware being as high as 6.4 and 7.3 bit/s/Hz/user for the FZF and the PZF precoding scheme respectively. At an IBO of 5 dB, the HWIs remain considerable as the systems suffers from an average loss of 1 bit/s/Hz/user for the FZF precoder or up to 1.5 for the PZF. The omnipresence of the distortions in the signal contribution is further demonstrated by the spread of the CDF, which is smaller as the hardware distortions gets stronger. The second aspect of Figure 1 focuses on the gain of performance offered by the PZF precoding scheme. On the ideal hardware case, the PZF scheme is more efficient in our deployment scenario, with a minimum gain of 15\% in SE. This is possible thanks to the large path loss between distant UEs and APs being compensated by the use of the MRT approach for weak users, whereas UEs in the vicinity of a given AP are still benefiting from the IUI cancellation offered by the zero-forcing applied to strong UEs.  In this case, we have optimized the grouping threshold $\nu_th = 99\%$ to our scenario with small increments of its value through simulations in order to maximize the achievable performance of the PZF scheme.

\begin{figure}[ht!] 
    \centering
    \includegraphics[width=0.85\linewidth]{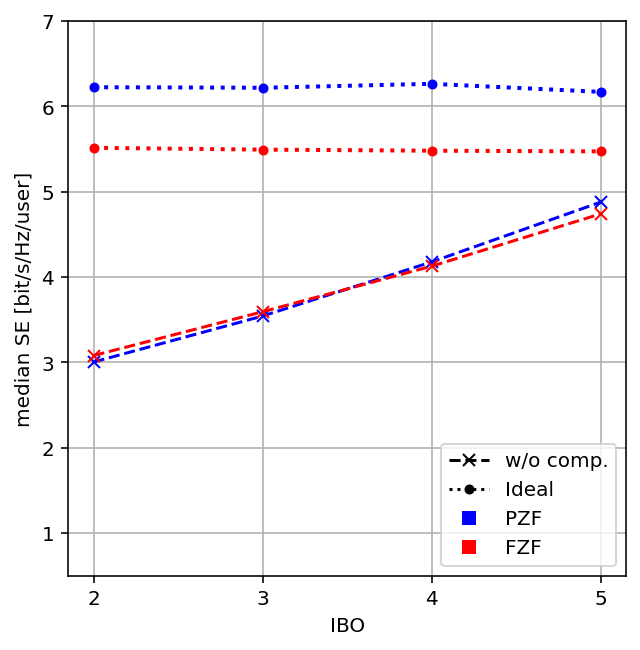}
    \caption{Median of the SE for the FZF and the PZF precoder ($\nu_{th}=99\%$) against IBO. $\eta^{DL}_{l,k}= 18.52 \mathrm{dBm}$} 
    \label{fig:Median_1}
\end{figure}

In Figure \ref{fig:Median_1}, the median SE [bit/s/Hz/user] against the IBO [dB] is given. It shows that, thanks to our normalization of the DL transmit power w.r.t the IBO, the displayed performance in the absence of HWI are independent to the IBO level used in the experiment. It allows to evaluate the efficiency of the hardware compensation algorithms studied in this work against a given level of distortions with no variations of the performance due to added efficiency of the PA or any other phenomenon dependent on the intrinsic characteristic of the considered PA. Here, we see that the PZF precoding fails to improve the median SE when the system is subject to HWI, even at IBO as high as 5 dB. Consequently, the saturation-induced distortions are the key limiting factor to improve DL system performance in the evaluated scenario. This further motivates the design of signal processing solutions aimed at lowering the impact of the PA on CF-mMIMO system.  

\subsection{Compensation of HWI : spectral efficiency}

\begin{figure}[ht!] 
    \centering
    \includegraphics[width=0.85\linewidth]{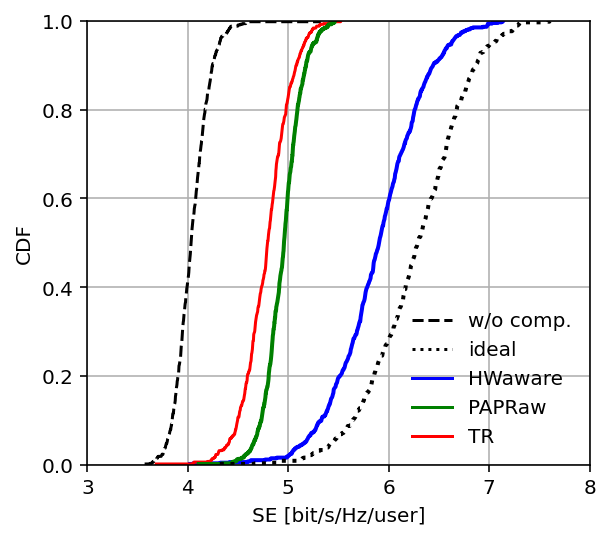}
    \caption{CDF of the SE : reference scenario with $M=8$ and IBO $= 4$ dB for $K=\tau_p=7$}
    \label{fig:CDF_2}
\end{figure}

On Figure \ref{fig:CDF_2}, we show the CDF of the SE with an IBO of 4 dB and 8 antennas per AP for the three studied solutions : Sequential hardware-aware precoding (HWaware), PAPR-aware precoding (PAPRaw) and Tone reservation (TR). The TR and PAPR-aware use the settings offering a suitable complexity to performance trade-off for this particular case, i.e, $N_{Tr} = 8$ with $N^{TR}_{it} = 15$ iterations, and the PAPR-aware precoder is constrained to $N^{PAPR}_{it} = 5$. It can be seen that the PAPR-aware algorithm is the most dependent on the ratio of user per antenna, as reducing the number of UEs to 5 greatly increased the reductions in hardware distortions. In contrast, the TR and HW-aware precoder offers similar performance w.r.t to their level of HWI compensation for both scenarios. The HW-aware precoder is greatly improving the performance, with a worst-case SE of 4.6 and a maximum of 7.1 bit/s/Hz/user compared to the 3.5 to 5 bit/s/Hz/user without compensation. Meanwhile, the “classical” method based on tone reservation is increasing the SE by 1 bit/s/Hz/user at most. When the number of UEs to be served is reduced (Fig.3.b), the difference between the PAPR-aware and HW-aware solutions and the ideal performance is sensibly lower, most notably at the higher end of the tail, with the PAPR-aware precoder going above the performance of HW-aware precoder. For the PAPR-aware approach, this is due to a better reduction of the PAPR for higher number of antenna as demonstrated in the PAPR reduction analysis section latter in this work. In contrast, the HW-aware precoder is only subjects to the mismatch between the noisy estimated channel and the actual propagation condition to the UEs as seen on (\ref{psi2}) for the strong set of users, while the weak users remains uncompensated with little impact on the total performance, especially for the worst cases in the lower end of the distribution.

\begin{figure*}[hbt!] 
    \centering
    \includegraphics[width=0.70\linewidth]{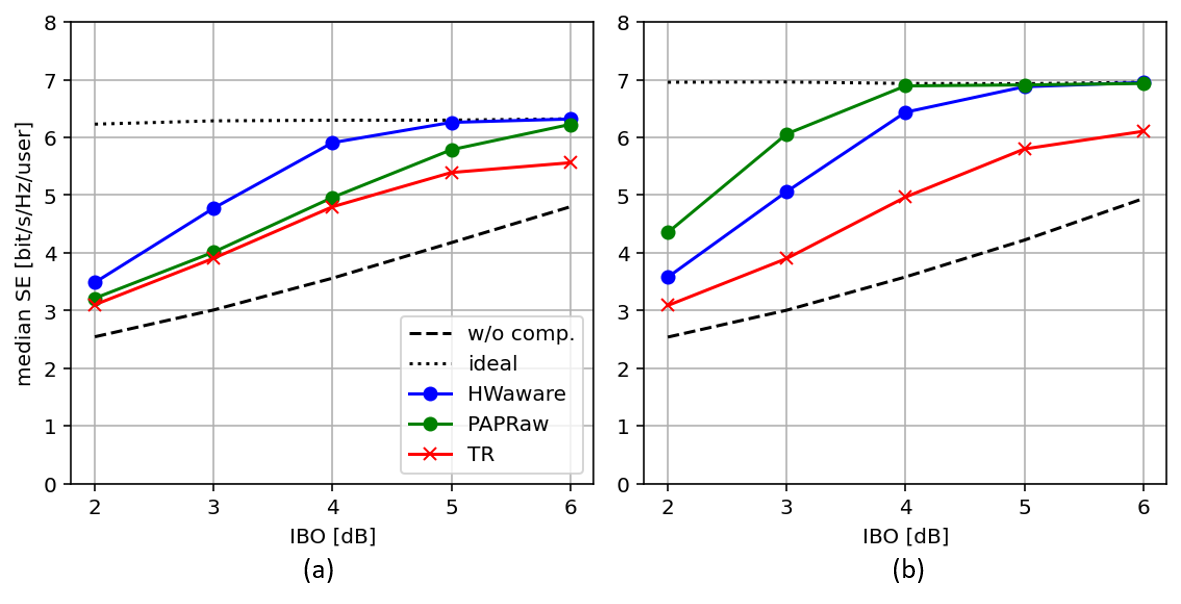}
    \caption{Median SE against IBO for $K=7$ and $PN_{iter} = 5$, $N_{Tr} = 8$ for : (a) $M=8$, (b)  $M=16$}
    \label{fig:Med_2}
\end{figure*}

\begin{figure*}[ht!] 
    \centering
    \includegraphics[width=0.70\linewidth]{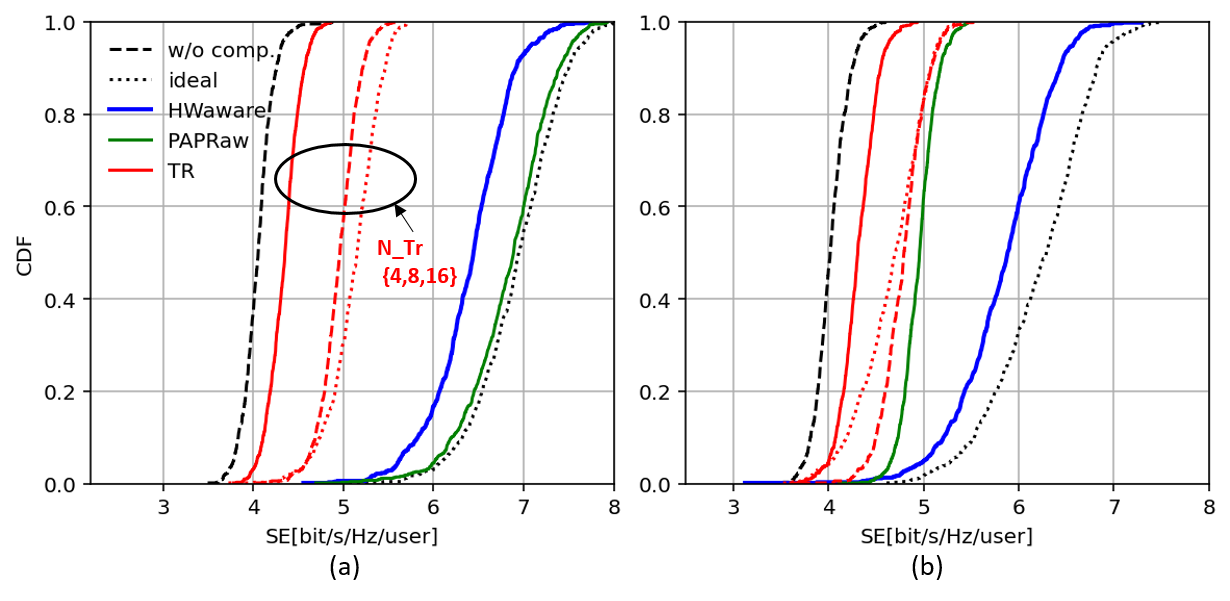}
    \caption{CDF of the SE for $IBO = 4 dB$ and $K=\tau_p=7$ and : (a) $M=16$, (b) $M=8$}
    \label{fig:CDF_3}
\end{figure*}

This is confirmed by Figure \ref{fig:Med_2}.a, where the median SE against the IBO is given for $M=8$ and $K=7$. The HW-aware precoder outperforms the other solutions for every value of IBO, with the PAPR-aware precoder being substantially better than the TR only for the highers value of IBOs ($\mathrm{IBO} > 4$ dB). Notably, the HW-aware solution coupled to the PZF precoder reaches ideal levels of performance as long as the IBO is kept above 5 dB, upgrading the SE by 2 bit/s/Hz/UE compared to the non-compensated scenario for this value of IBO. It is also seen that the TR is still performing correctly in terms of achieved median SE, with a notable gain of 1.3 bit/s/Hz/UE at an IBO of 4 dB, however, it is more limited in the potential performance benefits than the CF-mMIMO specific methods. Interestingly, the PAPR-aware solution outperforms the HW-aware equivalent for every levels of IBO for a scenario with an higher number of antennas ($M=16$, Figure \ref{fig:Med_2}.b).
It is further confirmed with the CDF of Figure \ref{fig:CDF_3}, where the performance with 16 antennas are almost ideal when the PA is operated with an IBO of 4 dB, and substantial gains are achieved for the worst distortions (lower end of the tail) with 8 antennas. Meanwhile, the TR provides a moderate improvement in PAPR reduction and SE independently of the number of antennas. Nevertheless, for the TR method, it comes at the cost of a section of the frequency band being unusable for service towards the users. It can be seen on Figure \ref{fig:CDF_2}.b, where the performance of TR with $N_{Tr} = 16$ is bellow those of $N_{Tr} = 8$ because the additional reduction of PAPR is not enough to compensate for the loss of a larger part of the bandwidth. The sequential HW-aware precoder performs steadily good independently of the the number of antennas, as the constraint on the performance gain w.r.t the non-compensated signal is mostly defined by the quality of the CSI, which is determined by the noise level.

\begin{figure}[h] 
    \centering
    \includegraphics[width=0.99\linewidth]{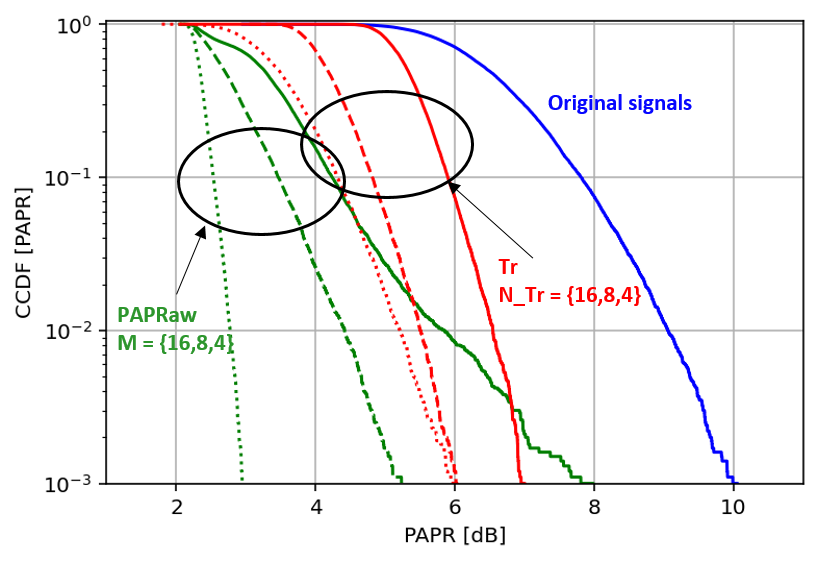}
    \caption{CCDF of the PAPR : PAPR-aware efficiency in PAPR reduction is proportional to the number of antennas; whiel TR is more efficient with a larger number of sub-carrier reserved for storing the displaced distortions}
    \label{fig:CCDF_PAPR}
\end{figure}

\subsection{PAPR reduction}

The effectiveness of PAPR-aware and TR techniques in reducing PAPR is evident when compared to the original PAPR distribution of the OFDM signals. This is illustrated in the Complementary Cumulative Distribution Function (CCDF) of PAPR shown in Figure \ref{fig:CCDF_PAPR}, where key parameters are adjusted to showcase the range of achievable performance for each method. For the PAPR-aware technic, the number of antennas of the system is varied from 16 to 4 per APs. Meanwhile, the TR methods relies on using a number of sub-carrier allocated to store the HWI and reduce the PAPR, here, we vary the number of reserved tone $N_{Tr} = {16;8;4}$ to be taken among the 64 sub-carriers of the system (25\%, 12.5\% and 6.25\% of the band respectively). Tone reservation achieves a significant reduction of the PAPR, most notably when $N_{TR} \geq 8$, with a 2 dB reduction at the top of the distribution while the worst-case are reduced to 6 dB as opposed to 10 dB for the original distribution.
For the PAPR-aware method, the key variable that contributes towards PAPR reduction is the number of antenna. This is due to the fact that the clipping noise is "stored" in the spatial plan towards angular direction that minimize the impact on the users, hence, the greater the degree of freedom, the better the ability of the system to reduce PAPR. While the original signals' PAPR is ranging between 5 to 8 dB in more than 9 out of 10 occurrences, PAPR-aware methods lowers it to down to 2dB at the top of the CCDF. This is a reduction by a factor 3 of the maximum PAPR exhibited when the PAPR-aware method is used with 16 antennas. Yet, the PAPR-aware algorithm demonstrates a clear dependency on the number of antennas per AP towards the bottom of the distribution. As the antenna count increases, the PAPR performance significantly improves. This highlights the effectiveness of the local PAPR-aware solution with the PZF grouping strategy to reduce the PAPR of the signals in CF-mMIMO networks. However, CF-mMIMO systems are expected to be deployed with a low number of antennas per AP to maintain cost effectiveness, which may results in this solution providing moderate performance gain for small APs. 

\subsection{Evaluation of complexity}

\begin{figure*}[h] 
    \centering
    \includegraphics[width=1.01\linewidth]{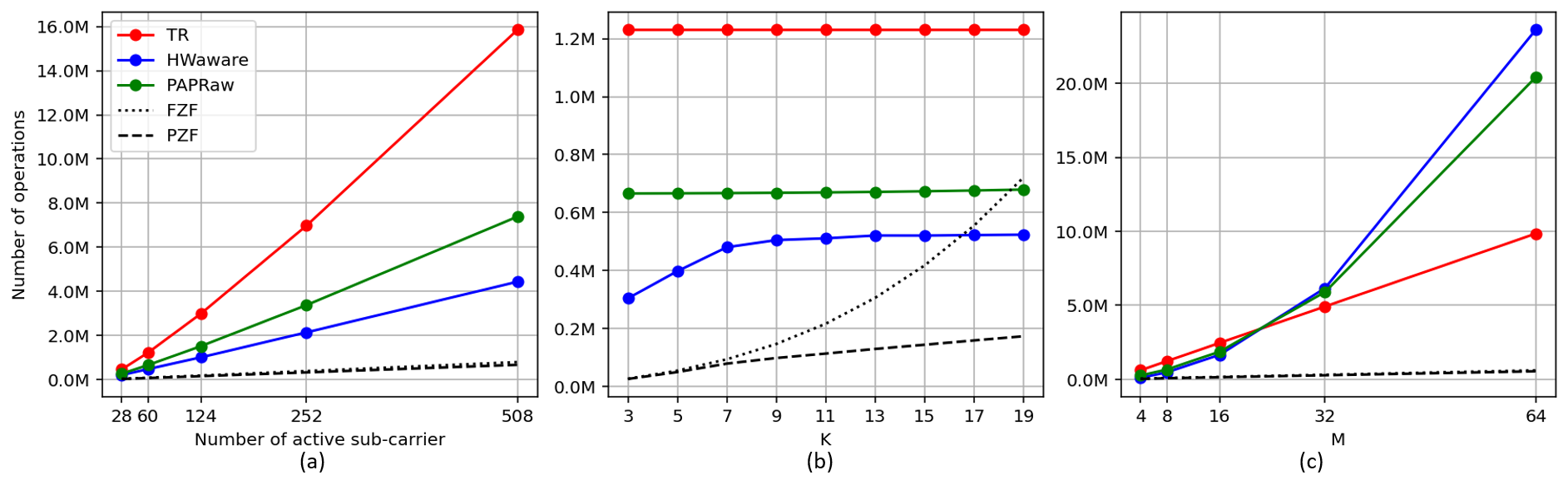}
    \caption{Average mean per-AP complexity (in millions [M] of complex multiplications over 1000 snapshot) of the FZF and PZF precoder and the three HWI compensation methods using the reference scenario ($K=\tau_p=7$, $|\Xi|=60$,$M=8$) as a base against : (a)  number of active subcarrier, (b) number of user, (c) number of antenna.}
    \label{fig:Complexity}
\end{figure*}

The complexity analysis for the three HWI compensation methods—Tone Reservation (TR), PAPR-aware precoding, and hardware-aware (HW-aware) precoding—is illustrated in Fig.~\ref{fig:Complexity}, highlighting their behavior under varying system parameters. Here, the complexity is averaged over 1000 snapshot where the size of the strong set of UE is determined according to the PZF grouping threshold and channel properties. The number of iteration of the TR and PAPR-aware algorithm remains at given in the previous section ($N^{PAPR}_{it} = 5$, $N^{TR}_{it} = 15$ as it has been tuned to offer optimal performance. When considering the number of active sub-carriers (Fig. \ref{fig:Complexity}.a), the TR algorithm demonstrates the steepest increase in complexity, primarily due to the extensive FFT/IFFT operations performed to move the clipped magnitudes in the reserved sub-carriers, making it highly sensitive to this parameter. In contrast, the HW-aware precoder maintains the lowest complexity, capped at $4$ millions of operations even with $N_{sc}\in|\Xi| = 508$, showcasing its suitability for scenarios with high number of sub-carriers. The resulting decrease of complexity of the HW-aware precoder approach compared to the PAPR-aware and the TR methods is respectively $40\%$ and $72\%$ with $508$ sub-carrier.

The scaling with the number of users ($K$) in Fig. \ref{fig:Complexity}.b reveals a distinct behavior across methods. While the TR and PAPR-aware precoders show negligible dependency on $K$, as the latter scales with $K$ only for the computation of $\mathbf{V}_0$, the HW-aware precoder shows a noticeable increase in complexity for smaller user loads ($K \in [3,9]$). Comparatively, in the one hand, the FZF precoder exhibits an exponential growth in complexity with increasing $K$, due to the much larger matrix operations required to perform zero-forcing. It even reaches higher complexity figure than the HW-aware and PAPR-aware methods when the number of users is high, which is  problematic considering the precoding scheme is computed only once per coherent channel intervals and therefore is inherently less computationally heavy. On the other hand, the HW-aware precoder instead reaches a plateau when $K=9$, probably due to the strong set of user having reached the maximum sustainable size for the channel contributions threshold   with that much users. At $K=19$, the HW-aware precoding requires 0.52 millions of complex multiplication, which is a reduction by respectively $23\%$ and $57\%$ compared to the PAPR-aware and the TR methods. It indicates its robustness for large user deployments thanks to relying on the PZF user grouping. Indeed, the PZF precoder shows a linear increase with the number of user but its complexity remains small even with large number of users : 0.19 millions of operation at $K=19$, when 0.72 millions are required for the FZF. 

The impact of the number of antennas per AP ( Fig. \ref{fig:Complexity}.c) introduces another dimension to the analysis. For larger values of $M$ (e.g., $M \in [32,64]$), the TR algorithm achieves the lowest complexity dues to not relying on matrix operation but rather on per-antenna clipping and FFT/IFFT manipulations. However, this behavior is not-so interesting for CF-mMIMO systems, which typically feature fewer antennas per AP. Both the HW-aware and PAPR-aware precoders exhibit a sharp increase in complexity with $M$, reaching $24$ millions of complex multiplications at $M = 64$. Despite this, they maintain reasonable complexity levels for realistic CF-mMIMO configurations ($M \in [4,16]$), ensuring their practical viability.

These results underline the HW-aware precoder's superior adaptability across varying system parameters, particularly in environments with a high number of sub-carriers or users. Although the TR algorithm benefits from lower complexity for high antenna counts, its dependence on FFT/IFFT operations limits its practicality for CF-mMIMO settings, while the PAPR-aware approach offers a reasonable complexity increase compared to the HW-aware, especially as it offers good performance even with a low number of iteration. Overall, the HW-aware approach strikes a favorable balance between computational efficiency and scalability. Additionally, we expect the sequential HW-aware precoder to totally erase the need for DPD-based linearization of the PAs as it directly compensate the distortions at the UEs, which suppress another layer of computational load on the signal processing as opposed to the other approaches.

\section{Conclusion}
In this study, we have conducted a comprehensive analysis of three hardware impairment compensation methods taken either from centralized massive MIMO literature (Tone Reservation) or
tailored for CF-mMIMO networks (PAPR-aware precoding), and a novel sequential hardware-aware sequential precoding scheme within the context of CF-mMIMO networks. Our findings indicate that the sequential HW-aware sequential precoding scheme offers high spectral efficiency performance even when the PA are operating near the saturation region. The complexity analysis underline the importance of considering computational scalability alongside performance gains when selecting an HWI compensation method. The HW-aware sequential precoding scheme's have shown its ability to maintain lower complexity across various configurations. Hence, digital post-distortion applied in an over-the air manner thanks to serial fronthaul topology is as a promising solution for next-generation CF-mMIMO networks. Future research should explore the ability of HW-aware precoding to deliver high performance on more complex hardware impairment models in order to evaluate the impact of the divergence between the pre-computed distortions and the real transfer function of the PAs on the performance of the solution. Furthermore, synchronization issues for such over-the-air based CF-mMIMO solution has to be considered.


%



\ifCLASSOPTIONcaptionsoff
  \newpage
\fi



\bibliographystyle{IEEEtran}
\bibliography{bibliography}
\end{document}